\documentclass[twocolumn]{aastex7}

\begin{document}
\title{The Extended Stellar Distribution in the Outskirts of the Ursa Minor Dwarf Spheroidal Galaxy}
\author[orcid=0000-0001-8239-4549,sname='Sato']{Kyosuke S. Sato}
\affiliation{Astronomical Science Program, The Graduate University for Advanced Studies, SOKENDAI, 2-21-1 Osawa, Mitaka, Tokyo 181-8588, Japan}
\affiliation{National Astronomical Observatory of Japan, 2-21-1 Osawa, Mitaka, Tokyo 181-8588, Japan}
\email[show]{kyosuke.sato@grad.nao.ac.jp} 

\author[orcid=0000-0002-7866-0514,sname='Okamoto']{Sakurako Okamoto}
\affiliation{National Astronomical Observatory of Japan, 2-21-1 Osawa, Mitaka, Tokyo 181-8588, Japan}
\affiliation{Astronomical Science Program, The Graduate University for Advanced Studies, SOKENDAI, 2-21-1 Osawa, Mitaka, Tokyo 181-8588, Japan}
\affiliation{Subaru Telescope, National Astronomical Observatory of Japan, 650 North A'ohoku Place, Hilo, HI 96720, USA}
\email{sakurako.okamoto@nao.ac.jp}

\author[orcid=0000-0001-7550-2281,sname='Yagi']{Masafumi Yagi}
\affiliation{National Astronomical Observatory of Japan, 2-21-1 Osawa, Mitaka, Tokyo 181-8588, Japan}
\affiliation{Department of Advanced Sciences, Faculty of Science and Engineering, Hosei University, 3-7-2 Kajino-cho, Koganei, Tokyo 184-8584, Japan}
\email{masafumi.yagi@nao.ac.jp}

\author[orcid=0000-0002-3852-6329,sname='Komiyama']{Yutaka Komiyama}
\affiliation{Department of Advanced Sciences, Faculty of Science and Engineering, Hosei University, 3-7-2 Kajino-cho, Koganei, Tokyo 184-8584, Japan}
\affiliation{National Astronomical Observatory of Japan, 2-21-1 Osawa, Mitaka, Tokyo 181-8588, Japan}
\email{komiyama@hosei.ac.jp}

\author[sname='Arimoto']{Nobuo Arimoto}
\affiliation{National Astronomical Observatory of Japan, 2-21-1 Osawa, Mitaka, Tokyo 181-8588, Japan}
\email{nobuo.arimoto@gmail.com}

\author[orcid=0000-0002-4013-1799,sname='Wyse']{Rosemary F.G. Wyse}
\affiliation{Department of Physics and Astronomy, Johns Hopkins University, Baltimore, MD 21218, USA}
\email{wyse@jhu.edu}

\author[orcid=0000-0001-6196-5162,sname='Evan']{Evan N. Kirby}
\affiliation{Department of Physics and Astronomy, University of Notre Dame, Notre Dame, IN 46556, USA}
\email{ekirby@nd.edu}

\author[orcid=0000-0002-9053-860X,sname='Chiba']{Masashi Chiba}
\affiliation{Astronomical Institute, Tohoku University, Aoba-ku, Sendai, Miyagi 980-8578, Japan}
\email{chiba@astr.tohoku.ac.jp}

\author[orcid=0000-0001-8239-4549,sname='Ogami']{Itsuki Ogami}
\affiliation{National Astronomical Observatory of Japan, 2-21-1 Osawa, Mitaka, Tokyo 181-8588, Japan}
\affiliation{The Institute of Statistical Mathematics, 10-3 Midoricho, Tachikawa, Tokyo 190-8562, Japan}
\affiliation{Department of Advanced Sciences, Faculty of Science and Engineering, Hosei University, 3-7-2 Kajino-cho, Koganei, Tokyo 184-8584, Japan}
\email{itsuki.ogami@nao.ac.jp}

\author[sname='Tanaka']{Mikito Tanaka}
\affiliation{Department of Advanced Sciences, Faculty of Science and Engineering, Hosei University, 3-7-2 Kajino-cho, Koganei, Tokyo 184-8584, Japan}
\email{mikito@hosei.ac.jp}

\begin{abstract}
	We discover an extended distribution of main-sequence (MS) stars along the minor axis of the Ursa Minor (UMi) dwarf spheroidal galaxy (dSph).
	This study is enabled by deep, wide Subaru/Hyper Suprime-Cam data, reaching photometric uncertainties below 0.1 mag at $g,i \sim 26$ mag.
	Color-magnitude diagrams along the major and minor axes reveal a clear excess of MS stars beyond the nominal tidal radius along the minor axis.
	To characterize this structure, we derive radial number density profiles in seven azimuthal directions and fit them with an exponential+power-law function to assess the symmetry of the extended component.
	The power-law slopes tend to be shallower toward the minor axis, though the symmetry remains inconclusive within 1$\sigma$ uncertainties.
	This may indicate that the extended component is preferentially distributed along the minor axis, and could be different from the previously suggested tidal features along the major axis.
	Comparing with simulations, we find that the fraction of stars beyond five effective radii is consistent with expectations from an intermediate mass ratio merger scenario with a stellar mass ratio around 6:1.
	While these findings provide new insights into the structural complexity and dynamical history of the UMi dSph, alternative mechanisms such as stellar or supernova feedback have also been proposed for extended stellar halos in dwarfs and cannot be ruled out.
\end{abstract}

\section{Introduction}\label{sec:intro}
In the $\Lambda$CDM cosmological model, structure formation in the Universe proceeds hierarchically, driven by the growth of initial dark matter density fluctuations (e.g., \citealp{1988ApJ...327..507F}).  
These fluctuations are amplified over time, leading to the assembly of cosmic structures from small to large scales, i.e., from dwarf galaxies to massive galaxies and galaxy clusters, following the shape of the initial power spectrum.  
Within this framework, dwarf galaxies are among the first systems to form, and represent the smallest building blocks in hierarchical clustering.  
Some of these dwarfs are thought to be the surviving relics of the earliest galaxies.  
Because the chemical abundances of their stars preserve information from the first billion years of cosmic history, dwarf galaxies serve as key probes for studying galaxy formation in the early Universe (e.g., \citealp{2015ARA&A..53..631F}).

Milky Way (MW) halo structure has been formed by accretion and merging of smaller stellar systems in a hierarchical structure formation scenario (e.g., \citealp{1978ApJ...225..357S,2001ApJ...558..666B, 2005ApJ...635..931B,2010MNRAS.406..744C}).
The remnants of past accreted systems are discovered as substructures in the MW halo (e.g., \citealp{2018Natur.563...85H, 2021ApJ...920...51M, 2022Natur.601...45M}).
Detecting and characterizing stellar halos in dwarf galaxies are crucial for understanding whether hierarchical merging processes were also active in low mass galaxies.
Several studies have reported extended stellar distributions around several Local Group dwarf galaxies, and are considered to be candidate of stellar haloes of dwarf galaxies \citep{2021NatAs...5..392C, 2024MNRAS.527.4209J, 2024ApJ...971..107O, 2021ApJ...923..218F}.  
\citet{2021NatAs...5..392C} suggested that the origin of the extended stellar distribution around the Tucana II ultra-faint dwarf (UFD) galaxy is due to a past dwarf-dwarf merger or stellar or supernovae feedback.  
That study found that the stars located at large distances from the galaxy center tend to exhibit lower metallicities, and no clear velocity gradient associated with tidal distortion is observed.
Moreover, the spatial distribution of these stars is oriented perpendicular to the orbital direction of Tucana II.
This extended stellar distribution not along the orbit is reproduced by some N-body simulations of dwarf-dwarf merger \citep{2021ApJ...914L..10T, 2022MNRAS.511.4044D, 2025A&A...694A..17Q}.

The Ursa Minor (UMi) dSph is a strong candidate for having a stellar halo despite being the lowest stellar-mass galaxy among the classical dSphs. 
\citet{2024MNRAS.527.4209J} investigated the stellar distributions in 60 MW dwarf galaxies using {\it Gaia} DR3 data, and found extended structure in nine dwarf galaxies, including the UMi dSph.  
\citet{2023MNRAS.525.2875S} investigated five member stars of the UMi dSph located beyond its nominal tidal radius and confirmed their membership through spectroscopic follow-up.
By examining the logarithmic derivative of the surface density profile, they identified a change in slope, named as a “kink” and a “break”, which they attributed to the effects of tidal forces. 
However, the nature of such extended structures remains observationally uncertain, as their conclusions are based on a limited number of red giant branch (RGB) stars.

Meanwhile, the extended structure of the UMi dSph could have been formed by a dwarf-dwarf merger.
\citet{2020MNRAS.495.3022P} reported two stellar populations in the UMi dSph that are distinct in metallicity, velocity dispersion, and spatial distribution.
The metal-poor population has a mean metallicity of $\rm [Fe/H] = -2.05 \pm 0.03$, a velocity dispersion of $\sigma_v = 4.9^{+0.8}_{-1.0}\ \rm km\ s^{-1}$, and an ellipticity of $\epsilon = 0.75 \pm 0.03$, while the extremely metal-poor population has $\rm [Fe/H] = -2.29^{+0.05}_{-0.06}$, $\sigma_v = 11.5^{+0.9}_{-0.8}\ \rm km\ s^{-1}$, and $\epsilon = 0.33^{+0.12}_{-0.09}$.
In addition, \citet{2020MNRAS.495.3022P} reported weak signatures of prolate rotation about the major axis, which is aligned with the orbital direction of the UMi dSph.
The star formation histories (SFHs) of these two populations recently derived by \citet{sato2025starformationchemicalevolution} show that they have different SFHs.
Moreover, the SFHs of both metal-poor and extremely metal-poor stars show radial variation but no clear age gradient.
These findings suggest that the complex structure results from the mixing of stellar populations originating from distinct stellar populations due to a recent dwarf-dwarf merger.

Although a recent merger event has been suggested for the UMi dSph \citep{2020MNRAS.495.3022P,sato2025starformationchemicalevolution}, clear structural signatures of such an event have not been identified from the spatial distribution of RGB stars.
Due to the limited number of RGB stars, previous studies have lacked the sensitivity needed to trace detailed structures, particularly at the outskirts of the UMi dSph.
At the distance of the UMi dSph ($69 \pm 4$ kpc, $m - M = 19.18 \pm 0.12$; \citealp{1999AJ....118..366M}), the limiting magnitude of previous observations is not deep enough to detect main-sequence (MS) stars.
To overcome this limitation, we use deep, wide-field imaging data from Subaru/Hyper Suprime-Cam (HSC) to investigate the spatial distribution of the UMi dSph, by using abundant MS stars.
This approach enables a more detailed exploration of how the extended structures were formed.
\section{Data and reduction}\label{sec:data}
	We use the same imaging data as presented in \citet{sato2025starformationchemicalevolution}, which were obtained with the Subaru/HSC with $g$- and $i$-bands (proposals ID:S15A-057 and S15A-OT08). 
	Details of the data reduction are given in that paper.
	The resulting photometric catalog reaches down to $g,i \sim 26$ mag with photometric errors less than 0.1 mag in both $g$- and $i$-bands, allowing for a reliable selection of MS stars in the outer regions of the UMi dSph.
	Galactic extinction is corrected for using the Galactic dust map by \citet{2011ApJ...737..103S}, which is the same procedure as \citet{sato2025starformationchemicalevolution}.
	Extinction corrected magnitude and color are indicated as $i_0$, and $(g-i)_0$.
	We apply the same star-galaxy separation criteria as in \citet{sato2025starformationchemicalevolution} to ensure consistency and reproducibility of the sample selection.

\section{Widely extended main-sequence stars}\label{sec:widelyextended}
\begin{figure*}[ht!]
	\begin{center}
	\includegraphics[width=18cm]{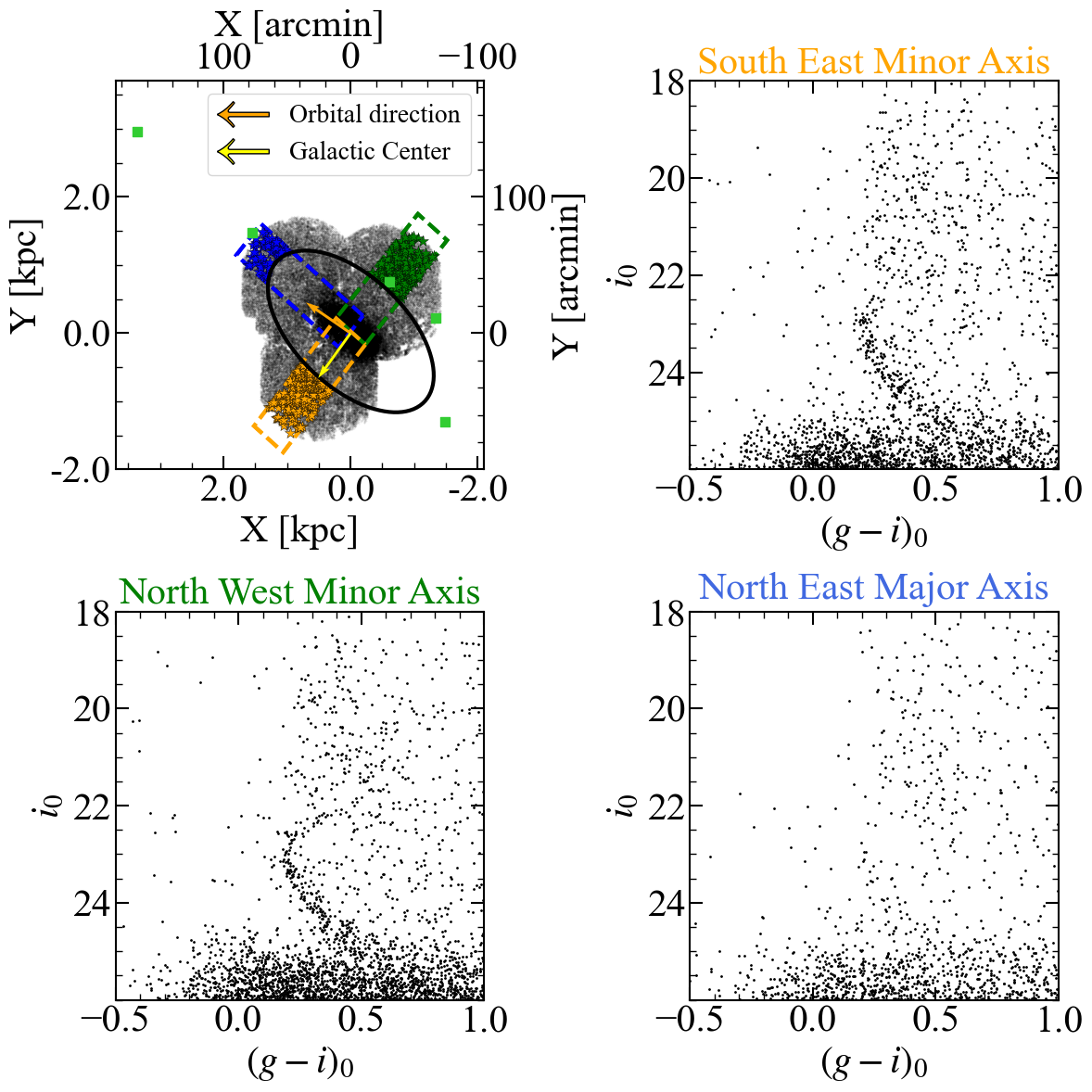}
	\end{center}
	\caption
	{
	The top-left panel shows the projected spatial distribution of all point sources with the footprint of the HSC dataset (black points). 
	The black ellipse indicates the nominal tidal radius of UMi dSph \citep{2010MNRAS.406.1220W}.
	The orange and yellow arrows point to the orbital direction and the MW center, respectively.
	The green squares represent halo star candidates of the UMi dSph that were spectroscopically confirmed by \citet{2023MNRAS.525.2875S}.
	The colored dashed boxes indicate the selected outer regions along the minor and major axes.
	The selected MS stars are represented by the colored stars.
	The top-right, bottom-left, and bottom-right panels show the CMDs of point sources located in the green (North West), orange (South East), and blue (North East) dashed regions, all of which lie outside the nominal tidal radius indicated by the black ellipse in the top-left panel.
	}\label{fig:spatial_cmd}
\end{figure*}

The top-left panel of Figure \ref{fig:spatial_cmd} displays the spatial distribution of the stars in our dataset (black points).
The green squares indicate the positions of the UMi dSph member stars located beyond the nominal tidal radius, as spectroscopically identified by \citet{2023MNRAS.525.2875S}.
The black ellipse denotes the nominal tidal radius of the UMi dSph ($77.9$ arcmin, $1.56$ kpc; \citealp{2010MNRAS.406.1220W}).
The dashed blue rectangle with size of $0.6\times2.1$ kpc marks the selection area at the major axis.
Similarly, the green and orange rectangular regions are selection areas at the minor axis, with the same dimensions as the blue one.
The orange and yellow arrows indicate the directions of the orbital motion of the UMi dSph and the center of the MW, respectively.
Notably, the spectroscopically confirmed stars are preferentially distributed at the major axis of the UMi dSph, which aligns with the direction of its orbital direction.
As the galaxy has completed multiple orbits around the MW and experiences repeated pericentric passages \citep{2020ApJ...905..109M}, it is thought to have undergone tidal stripping, leading to the formation of a tidally extended stellar structure in the outer regions \citep{2008ApJ...673..226P}.

The bottom-left, top-right, and bottom-right panels of Figure \ref{fig:spatial_cmd} present the color-magnitude diagrams (CMDs) for two regions along the minor axis and one along the major axis, specifically selected beyond the nominal tidal radius.
A clear distribution of MS stars is visible along the minor axis, whereas it is less apparent along the major axis.
To confirm the effect of the difference between the area of the major and minor axes, we search for the number of MS stars over $15$ arcmin by the nominal tidal radius in each direction.
We find that the selected areas at the South East (SE) and North West (NW) minor axes have 113 and 141 MS stars, respectively, while the North East major axis has 47 MS stars.
Even if the area is set as equal, the number of MS stars in the area along the major axis is smaller than that of the minor axes, even when the Poisson error is taken into account.
By comparing the stellar numbers along the SE and NW minor axes, we find that the NW region contains a larger number of stars.
When the area is expanded to $25$ arcmin for both minor-axis regions, the numbers are 173 and 186 for the SE and NW, respectively.
Thus, the apparent excess in the NW may be primarily associated with the inner regions of the UMi dSph.
\citet{2003AJ....125.1352P} reported the presence of an S-shaped structure in the UMi dSph.
The NW region lies precisely along this S-shaped curve, and it is likely to contain a large fraction of stars belonging to the main body of the UMi dSph.
This result suggests that MS stars are distributed beyond the nominal tidal radius along the minor axis.
While extended structures at the major axis have previously been identified by \citet{2023MNRAS.525.2875S}, the extended distribution observed at the minor axis in this study may have a different origin.


\section{Radial number density profile by directions}

\subsection{The shape of extended structure}\label{sec:shapeofextended}
\begin{figure*}[ht!]
	\begin{center}
	\includegraphics[width=14cm]{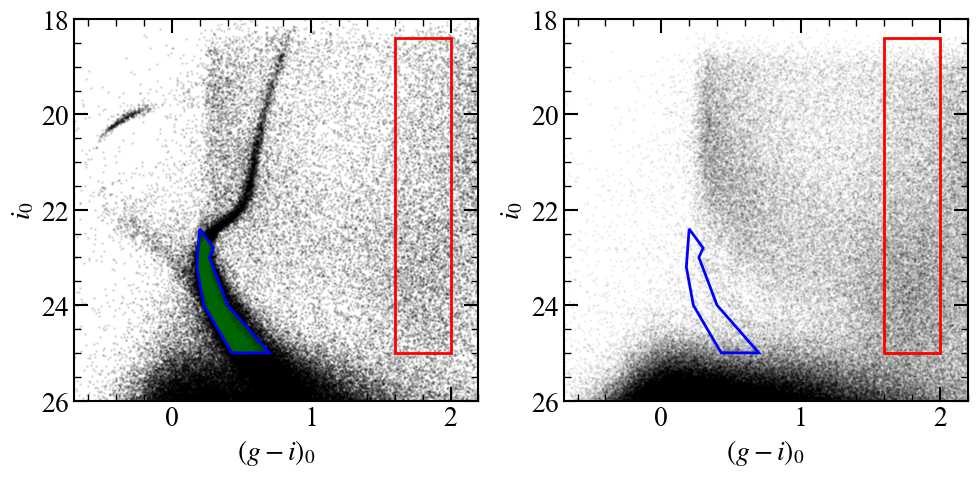}
	\end{center}
	\caption
	{
	The CMD of the UMi dSph (left) and control field (right).
	The blue rectangle is set to select the MS stars.
	Green points indicated in the left panel are selected member MS stars.
	The red rectangle in left panel is used to derive the normalized number density profile of contaminants.
	The red rectangle in the right panel is used to normalize the number of contaminants, selected within the blue polygon in the right panel as mimics of UMi member MS stars, by comparing it to the number of selected stars within the red rectangle in the left panel.
	}\label{fig:CMD}
\end{figure*}

In order to investigate the detailed structure at the outskirts of the UMi dSph, We focus on MS stars in this section.
The left panel of Figure \ref{fig:CMD} displays the point sources within the entire observed region, which includes UMi stars, as well as foreground and background contaminations.
The color and magnitude range of the RGB is heavily contaminated by foreground MW stars.
In contrast, the MS stars of the UMi dSph appear bluer than the foreground stars because the metallicity of the UMi member stars is lower than that of the MW foreground stars, resulting in less overlap in the CMD.
Furthermore, given the density law of the MW halo, the foreground contamination should be smaller at the distance of the UMi dSph ($i_0\sim 22$).
These factors enable us to select member stars with relatively low contamination even without relying on kinematic information.

To minimize contamination, we define a blue polygonal region in the left panel of Figure \ref{fig:CMD} to select member MS candidates.
Since background galaxies tend to populate the faint end ($i_0 > 25.0$; \citealp{2019ApJ...884..128O}), we limit our MS selection to $i_0 \leq 25.0$.
This selection is consistently applied to CMDs constructed in three directions at the major and minor axes (see top-left panel of Figure \ref{fig:spatial_cmd}).

To estimate the contamination of the selected member MS stars, we set the area ($240^{\circ} \leq \rm{R.A.} \leq 246^{\circ}, +42.5^{\circ} \leq\rm{Dec.} \leq +44.5^{\circ}$) in the HSC-SSP DR3 \citep{2022PASJ...74..247A} as the control field at similar Galactic latitude to the UMi dSph.
The CMD of selected control field is indicated in the right panel of Figure \ref{fig:CMD}.
We count the total number of point sources within the same MS selection polygon (blue polygon) in the control field. 
The contamination at the UMi field is estimated by scaling the number by the ratio of the point sources within the red rectangle shown in Figure \ref{fig:CMD} between the UMi field and the control field.

\begin{figure*}[h]
    \begin{center}
    \includegraphics[width=16cm]{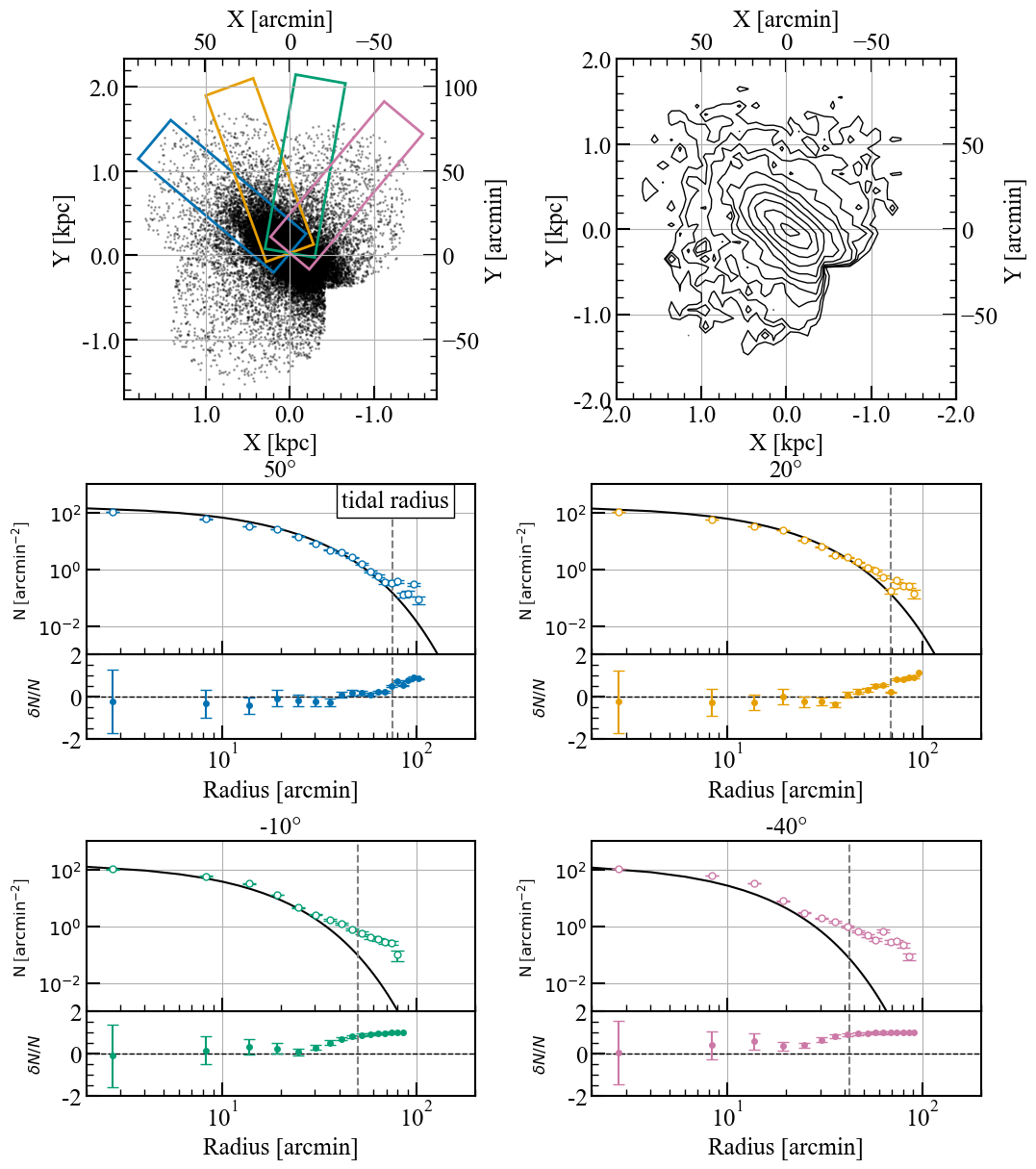}
    \end{center}
    \caption
    {
    The top-left panel shows the spatial distribution of stars, with regions divided by azimuthal angles for $50^\circ$ (blue), $20^\circ$ (yellow), $-10 ^\circ$ (green), and $-40^\circ$ (purple).
	The top-right panel presents a contour map of main-sequence (MS) stars selected using the CMD method described in Section \ref{sec:shapeofextended}.
	Black contours show the surface number density of MS stars, derived from a weighted 2D histogram. Each bin corresponds to an area of $100 \times 100 \, \mathrm{pc}^2 = 10{,}000 \, \mathrm{pc}^2$, and the contours are logarithmically spaced from $N_\ast = 10$ to $10^4$ stars per bin, corresponding to approximately $1.0 \times 10^{-3}$ to $1.0 \, \mathrm{pc}^{-2}$.
	The upper sub-panels in the middle and bottom rows compare the observed radial profiles with the exponential model (black line) representing the stellar distribution of the UMi main body.
	The lower sub-panels in each of these panels show the residuals.
    }\label{fig:azimunthal}
\end{figure*}

In Figure \ref{fig:azimunthal}, we select four azimuthal directions at position angles of $-40^\circ$, $-10^\circ$, $20^\circ$, and $50^\circ$ measured from north through east (i.e., from the direction where the position angle is defined $0^{\circ}$ points to the north and $+90^{\circ}$ to the east).
The $50^\circ$ and $-40^\circ$ direction correspond to the major and minor axis directions, respectively.
The middle and bottom four panels compare the observed radial profiles with the exponential profile predicted from the elliptical shape of the UMi main body ($\rm{P.A.}= 49.97\ \rm{deg}$, and $\epsilon=0.46$; \citealp{sato2025starformationchemicalevolution}).
We count the number of selected MS stars within binned regions in projected coordinates.
The bin width is set as $\sim5.15$ arcmin ($100$ pc).
We consider errors in both the radial profile of MS stars and the contamination profile.
\cite{2003AJ....125.1352P} discovered the S-shape morphology of the UMi dSph.
They suggest that this structure evolved by the tidal influence of the Milky Way.
However, this S-shape does not appear in the distribution of the MS stars (see top-right panel of Figure \ref{fig:azimunthal}).

We confirm that, in all four regions, the observed profiles exceed the predicted exponential profile, which is based on the values from \cite{sato2025starformationchemicalevolution}, at the outskirts.
The excess of the observed radial profiles over the single exponential profile at various directions indicates that the extended structure would be present in all directions.


\subsection{Model fitting}\label{sec:fitting}

\begin{figure*}[ht!]
	\begin{center}
	\includegraphics[width=18.1cm]{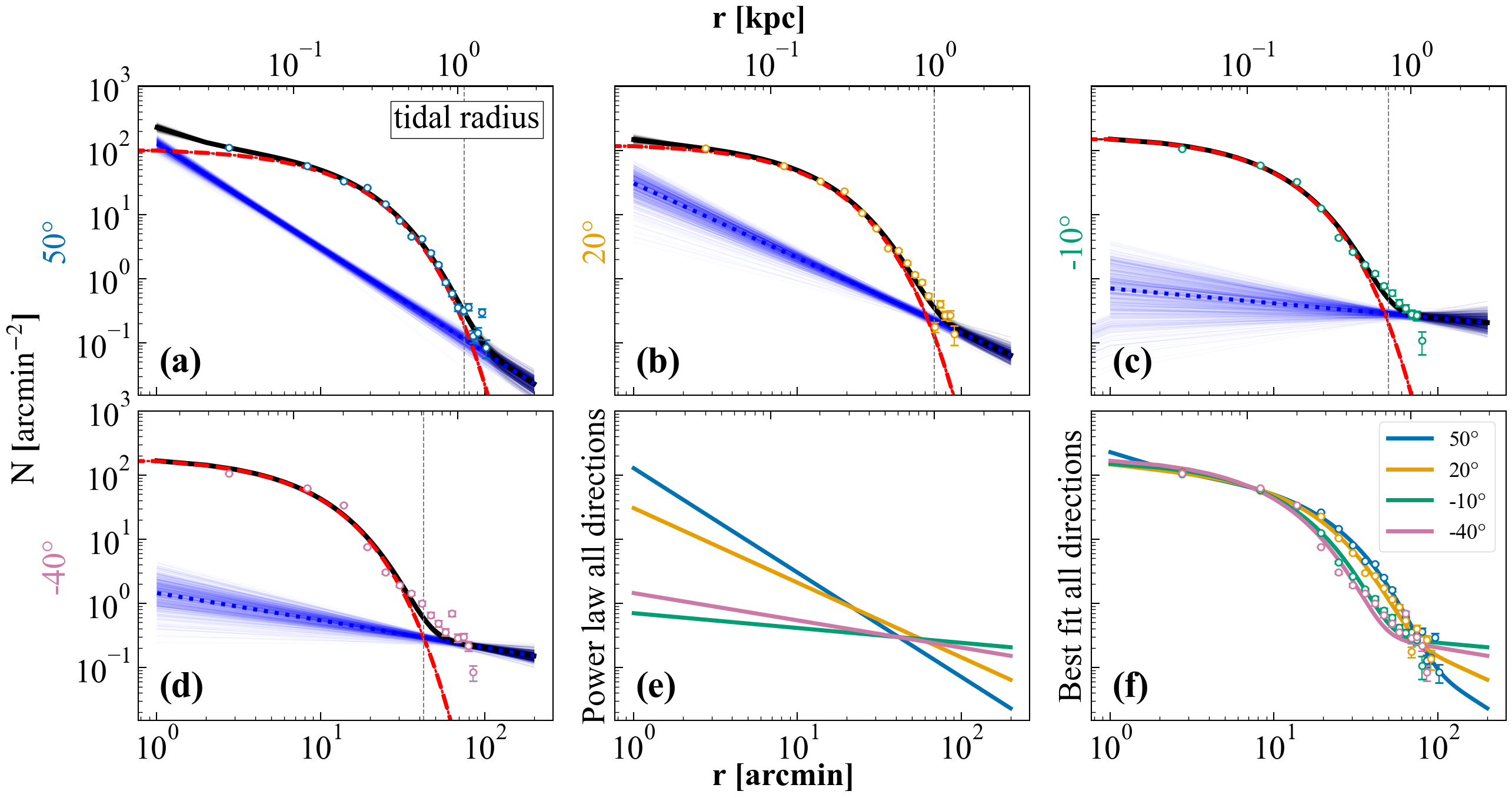}
	\end{center}
	\caption
	{
		Radial number density profiles and fitting results of MS stars in 4 directions. 
		Colored circles with error bars are the radial number density profiles of stars in each direction, with errors. 
		The solid black lines indicate the best-fit results obtained from MCMC fitting. 
		The red dashed and blue dotted lines correspond to the exponential and power-law models, respectively. 
		For each model, the thick line shows the median parameter values from the MCMC samples, while the thin lines represent the profiles generated by each of the 100 MCMC walkers. 
		The gray dotted line marks the nominal tidal radius corresponding to the black ellipse shown in Figure \ref{fig:spatial_cmd}.
		The panels (e) and (f) compare the power-law and best-fit profiles across four directions, with colored lines and open circles indicating the profiles and radial number density profiles, respectively.
	}\label{fig:radialprofile}
\end{figure*}

To investigate slope and stellar counts of this extended component, we fit a double component model of exponential and power-law profiles as
\begin{equation}
	\label{eq:model}
	\Sigma(r) = \Sigma_{\rm{UMi},0} \exp\left(-\frac{r}{r_{s}}\right)+\Sigma_{\rm{ext},1} \left(\frac{r}{\rm{arcmin}}\right)^{\alpha},
\end{equation}
where $r_s$ is the scale length, $\alpha$ is the projected power-law index which characterizes the extended structure of the UMi dSph, and the $\Sigma_{\rm{UMi},0}$, $\Sigma_{\rm{ext},1}$ are the surface number densities for the main body of the UMi dSph at the center and extended structure at $r=1$ arcmin, respectively.
Since the power-law profile is infinite at $r=0$ when $\alpha<0$, we need a normalization point somewhere at $r>0$.
$r=1$ arcmin is adopted arbitrarily, while it only affects the $\Sigma_{\rm{ext},1}$ value.
We fit the profile by the Markov Chain Monte Carlo (MCMC) using Python EMCEE module \citep{2013PASP..125..306F}. 
We use the uniform prior distribution for $\Sigma_{\rm{UMi},0},r_s,\Sigma_{\rm{ext},1}$, and $\alpha$.
The thick black lines in Figure \ref{fig:radialprofile}(a)--(d) are results of MCMC profile fitting for each of the four directions.
The red and blue lines are the estimated exponential and power-law profiles, respectively.
The parameter of exponential profiles converge tightly in all directions, making the individual walker lines (thin red lines) nearly indistinguishable.
The estimated parameters are presented in Table \ref{tab:parameters}.

The results indicate that number density profiles of all directions prefer the double component than single exponential profile.
In Figure \ref{fig:radialprofile}(e), we compare the slope of power-law profiles for each of the four directions.
We find that the slope becomes shallower as the direction approaches the minor axis, while the $-10^\circ$ region is steeper than the $-40^\circ$ region.
Nevertheless, the slopes in the $-10^\circ$ and $-40^\circ$ directions have similar parameters within 1$\sigma$ uncertainties.
The results suggest that the extended structure is more prominent along the minor axis than along the major axis, indicating a possible elongation in the minor-axis direction.
However, the number of data points beyond the nominal tidal radius is rather limited in the major-axis direction, which may lead to less reliable parameter estimates there.
As discussed in Sec. \ref{sec:widelyextended}, even when comparing regions that have the same area, the number of MS stars is smaller along the major axis.
This trend is consistent with the shallower slopes obtained along the minor axis.
Although this suggests a possible alignment of the extended structure with the minor axis, the current data are not sufficient to conclude this definitively, and wider coverage is required to confirm this.
To further examine the structural symmetry, we also performed the same fitting along the $80^\circ$, $110^\circ$, and $140^\circ$ directions, and the estimated parameter values are summarized in Table~\ref{tab:parameters}.

For regions symmetric with respect to the major axis (e.g., $-10^\circ$ and $110^\circ$), the scale radius $r_{s}$ is comparable.
The power-law parameters differ, with regions at $>50^\circ$ tending to show a steeper slope $\alpha$ and a larger surface number density $\Sigma_{\rm ext,1}$.
Along the minor axis ($-40^\circ$ and $140^\circ$), the power-law slopes agree within the 1$\sigma$ uncertainties.
These trends may indicate that only the minor axis shows symmetry within 1$\sigma$, while the other regions do not.
\citet{2013MNRAS.433..878L} found that the outer density slope of tidal tails is close to $-2$ in 3D, except during pericenter passages. 
When transforming from the 3D coordinate system $r = (x,y,z)$ to the projected 2D coordinate $R = (x,y)$, a power-law density profile $\rho(r) = \rho_{0}\, r^{-\alpha}\ (1<\alpha)$ becomes a surface density profile $\Sigma(R) \propto R^{-(\alpha - 1)}$ after integration along the line of sight ($z$ axis). 
Therefore, our measured slopes are comparable to the results of \citet{2013MNRAS.433..878L}, supporting the interpretation that the extended structure we detect is related to tidally stripped structure.

\begin{table*}[ht!]
	\caption{Estimated parameters.}\label{tab:parameters}
\centering
	\footnotesize
	\renewcommand{\arraystretch}{1.5} 
	\begin{tabular*}{15cm}{@{\extracolsep{\fill}}ccccc}
		\hline
		Direction & $\Sigma_{\rm{UMi},0}\ \rm{[N\ arcmin^{-2}]}$ & $r_{s}$ & $\Sigma_{\rm{ext},1}\ \rm{[N\ arcmin^{-2}]}$ & $\alpha $ \\
		\hline
		$-40^\circ$ (Minor Axis) &$191.96^{+2.14}_{-2.11}$ & $6.55\pm0.04$ & $1.46^{+0.86}_{-0.58}$ & $-0.43^{+0.12}_{-0.11}$\\
		$-10^\circ$ &$169.89^{+1.96}_{-1.92}$ & $7.39\pm0.05$ & $0.7^{+0.75}_{-0.40}$ & $-0.23^{+0.20}_{-0.17}$\\
		$20^\circ$ & $128.29^{+2.68}_{-2.58}$ & $9.89\pm0.09$ & $30.53^{+12.06}_{-11.15}$ & $-1.16^{+0.11}_{-0.09}$\\
		$50^\circ$ Major Axis& $107.73^{+1.64}_{-1.60}$ & $11.80\pm0.11$ & $128.18^{+16.71}_{-15.74}$ & $-1.63^{+0.05}_{-0.06}$\\
		$80^\circ$ & $128.73^{+2.15}_{-2.09}$ & $10.47\pm0.10$ & $53.57^{+18.07}_{-15.41}$ & $-1.58^{+0.11}_{-0.12}$\\
		$110^\circ$ & $186.13^{+2.15}_{-2.18}$ & 7.37$\pm0.05$ & $4.31^{+2.53}_{-1.80}$ & $-0.72^{+0.13}_{-0.11}$\\
		$140^\circ$ (Minor Axis)& $212.54^{+2.33}_{-2.36}$ & 6.65$\pm0.04$ & $2.97^{+1.48}_{-1.10}$ & $-0.58^{+0.11}_{-0.10}$\\
		\hline
	\end{tabular*}
\end{table*}

\section{The origin of extended structure}\label{sec:discussion}
The origin of extended stellar structures in dwarf galaxies remains a topic of active debate.
Some of MW dSphs and UFDs exhibit extended stellar distribution well beyond their nominal tidal radii.
Such structures may result from tidal distortion by the MW, whereby tidal forces strip stars from the galaxy, especially during close pericentric passages \citep{2008ApJ...673..226P, 2013MNRAS.433..878L, 2023MNRAS.525.2875S}.
Alternatively, internal processes such as dwarf-dwarf mergers may spread stars out to large distances from the galaxy center \citep{2021NatAs...5..392C, 2022MNRAS.511.4044D, 2025A&A...694A..17Q}.
While the extended stellar distribution along the orbital direction can be interpreted as a result of tidal distortion, the stellar distribution perpendicular to the orbital path found in this study may not be fully explained by tidal forces alone, suggesting the involvement of additional dynamical processes.

\cite{2013MNRAS.433..878L} presented N-body simulation results on the tidal tail of MW dwarf galaxies of different orbits. 
Their Fig. 8 shows how the angle between the tail and the MW direction viewed from the MW center changes with time.
Among the models, O5 is comparable to UMi dSph, as the ratio of apocentric and pericentric distances of the UMi dSph has been estimated as $r_{\rm{Apo}}/r_{\rm{Peri}}\approx2.2$ (assuming the heavy MW potential; \citealp{2022A&A...657A..54B}), while that of model O5 is $\approx2.5$.
According to \citet{2020ApJ...905..109M}, the UMi dSph is currently falling from the apocenter. 
Fig. 8 of \cite{2013MNRAS.433..878L} suggests that UMi would be presently in a phase where the angle of the tidal tail to the MW center ($\theta$) is expected to be relatively smaller, $10^\circ\lesssim\theta\lesssim30^\circ$.
We calculated how the tidal tail would appear when projected onto celestial coordinates.
As a result, the tidal tail is observed with $50^\circ$ -- $80^\circ$ offset to the minor axis of the UMi dSph.
Therefore, at this stage, we cannot conclude that this structure is indeed a tidal tail.

According to the cosmological simulations studies, merger events are known to contribute significantly to the formation of stellar haloes in dwarf galaxies.
According to the cosmological $N$-body simulations by \citet{2022MNRAS.511.4044D}, which model dwarf-mass galaxies with dark matter halo masses of $M_{\rm DM} \sim 10^{10}\,M_{\odot}$, minor and intermediate dwarf--dwarf mergers are particularly effective in producing extended stellar halo structures.
Minor mergers typically correspond to dark matter mass ratios of $10\!:\!1$ (with stellar mass ratios of $\sim 80\!:\!1$), while intermediate mergers have dark matter mass ratios of $5\!:\!1$ (corresponding to stellar mass ratios of $\sim 20\!:\!1$).  
In \citet{2025A&A...694A..17Q}, the $f_5$ value, defined as the stellar mass ratio between the entire system and the region beyond five times the effective radius, is discussed as a useful constraint on the mass of merger progenitors.
We estimate $\log f_5$ from the exponential+power-law profiles assuming the outer edge of the UMi dSph of $150 \leq r_{\rm edge} \leq 200\ \mathrm{arcmin}$. 
By comparing the result, $-0.95 < \log f_5 < -0.67$, to the Figure 8 of \citet{2025A&A...694A..17Q}, the mass ratio of progenitors would be $3\lesssim \mathrm{M1/M2}\lesssim10$.
In addition, assuming that the multiple stellar populations in the UMi dSph originated from a merger event, the progenitor masses of the metal-poor ([Fe/H] $= -2.05$) and extremely metal-poor ([Fe/H] $= -2.29$) components are estimated to be $6.3 \times 10^4\ M_\odot$ and $1.0 \times 10^4\ M_\odot$, respectively, based on the mass--metallicity relation (Equation~4 of \citealp{2013ApJ...779..102K}) and thus $M1_{\star}/M2_{\star}=6.3$.
These findings suggest that the characteristics observed in the UMi dSph, such as the presence of multiple stellar populations and an extended structure, are sufficient to account for the formation of its own stellar halo.

Another possible sign of the past merger event is a weak feature of prolate rotation ($A_{\mathrm{rot}} = 1.8 \pm 0.8\ \mathrm {km\ s^{-1}}$) in the UMi dSph \citep{2020MNRAS.495.3022P}.
Prolate rotation has also been observed in other Local Group dwarf galaxies, such as Andromeda II, a satellite of M31 \citep{2012ApJ...758..124H}, the Phoenix dwarf galaxy \citep{2017MNRAS.466.2006K}, and NGC6822 \citep{2020ApJ...903...10B}.
These rotations are likely the result of a past merger event \citep{2015ApJ...813...10E}.
In particular, in the case of Phoenix, the peculiar spatial distribution of young stars is aligned with its prolate rotation axis, suggesting that this configuration may be a remnant of a past dwarf-dwarf merger event \citep{2017MNRAS.466.2006K}.

\citet{2020MNRAS.495.3022P} also reported a difference in ellipticity between the metal-poor and extremely metal-poor populations in the UMi dSph.
The extremely metal-poor population has a more spherical shape ($\epsilon=0.33^{+0.12}_{-0.09}$) compared to the metal-poor population ($\epsilon=0.75\pm0.03$).
This implies that the extremely metal-poor stars are more evenly distributed, while the metal-poor population is more elongated along the major axis.
Interestingly, the direction of the extended stellar distribution found in this study is in agreement with the velocity peak of the prolate rotation, which is similar to the Phoenix case.
This morphological feature may also be interpreted as a remnant of a past merger event.
However, we do not yet have definitive evidence to distinguish whether the extended structure of UMi was formed through a merger event or as a result of tidal distortion or feedback from stellar or supernova feedback.

\section{Conclusions}\label{sec:concl}
In this work, we investigated the detailed structure of the UMi dSph using the abundant MS stars.
We derived CMDs in three directions of the UMi dSph at its major and minor axes, and confirmed that MS stars beyond the nominal tidal radius are particularly prominent at the minor axis.

To investigate the properties of extended structure, we derived radial number density profiles for the seven azimuthal directions in the projected coordinate system.
We found that the extended structure exists in all directions.
Then, we also fitted the double component profile model to estimate the symmetry of this extended structure.
We found that the slope of power-law profiles tend to be shallow toward the minor axis.
The extended structure is more prominent along the minor axis.
While tidal distortion may have contributed to the formation of this extended structure, we consider that a past merger event could also be a plausible origin.  
Specifically, the computed $f_5$ value, the stellar mass ratio between the entire system and the region beyond five times the effective radius, is consistent with that expected from an intermediate mass-ratio merger.  
Moreover, the stellar mass ratio inferred from the mass-metallicity relation supports a similar merger scenario.
To confirm the existence of this outskirts structure, photometric studies of the more outer regions and chemodynamical analyses are essential.

\begin{acknowledgments}
This work was supported by JSPS KAKENHI Grant Numbers JP18H05875, JP20K04031, JP20H05855, JP25K01047, JP24K00669.
This work was supported by JST SPRING, Japan Grant Number JPMJSP2104.
RFGW acknowledges support through the generosity of Eric and Wendy Schmidt, by recommendation of the Schmidt Futures program.

The Hyper Suprime-Cam (HSC) collaboration includes the astronomical communities of Japan and Taiwan, and Princeton University. 
The HSC instrumentation and software were developed by the National Astronomical Observatory of Japan (NAOJ), the Kavli Institute for the Physics and Mathematics of the Universe (Kavli IPMU), the University of Tokyo, the High Energy Accelerator Research Organization (KEK), the
Academia Sinica Institute for Astronomy and Astrophysics in Taiwan (ASIAA), and Princeton University. 
Data analysis was carried out on the large-scale data analysis system co-operated by the Astronomy Data Center (ADC) and Subaru Telescope, NAOJ. 
This research is based on data collected at the Subaru Telescope.
We are honored and grateful for the opportunity of observing the Universe from Maunakea, which has the cultural, historical, and natural significance in Hawaii.
The data are obtained from SMOKA, which is operated by the NAOJ/ADC.

The Pan-STARRS1 Surveys (PS1) have been made possible through contributions of the Institute for Astronomy, the University of Hawaii, the Pan-STARRS Project Office, the Max-Planck Society and its participating institutes, the Max Planck Institute for Astronomy, Heidelberg and the Max Planck Institute for Extraterrestrial Physics, Garching, The Johns Hopkins University, Durham University, the University of Edinburgh, Queen's University Belfast, the Harvard-Smithsonian Center for Astrophysics, the Las Cumbres Observatory Global Telescope Network Incorporated, the National Central University of Taiwan, the Space Telescope Science Institute, the National Aeronautics and Space Administration under Grant No. NNX08AR22G issued through the Planetary Science Division of the NASA Science Mission Directorate, the National Science Foundation under Grant No. AST-1238877, the University of Maryland, and Eotvos Lorand University (ELTE).

The authors would like to thank the referee for constructive comments, which helped improve the quality of this paper. 
\end{acknowledgments}
\bibliography{ref_umi_halo_rev.bib}

\begin{thebibliography}{}
\expandafter\ifx\csname natexlab\endcsname\relax\def\natexlab#1{#1}\fi
\providecommand{\url}[1]{\href{#1}{#1}}
\providecommand{\dodoi}[1]{doi:~\href{http://doi.org/#1}{\nolinkurl{#1}}}
\providecommand{\doeprint}[1]{\href{http://ascl.net/#1}{\nolinkurl{http://ascl.net/#1}}}
\providecommand{\doarXiv}[1]{\href{https://arxiv.org/abs/#1}{\nolinkurl{https://arxiv.org/abs/#1}}}

\bibitem[{H. {Aihara} {et~al.}(2022){Aihara}, {AlSayyad}, {Ando}, {Armstrong}, {Bosch}, {Egami}, {Furusawa}, {Furusawa}, {Harasawa}, {Harikane}, {Hsieh}, {Ikeda}, {Ito}, {Iwata}, {Kodama}, {Koike}, {Kokubo}, {Komiyama}, {Li}, {Liang}, {Lin}, {Lupton}, {Lust}, {MacArthur}, {Mawatari}, {Mineo}, {Miyatake}, {Miyazaki}, {More}, {Morishima}, {Murayama}, {Nakajima}, {Nakata}, {Nishizawa}, {Oguri}, {Okabe}, {Okura}, {Ono}, {Osato}, {Ouchi}, {Pan}, {Plazas Malag{\'o}n}, {Price}, {Reed}, {Rykoff}, {Shibuya}, {Simunovic}, {Strauss}, {Sugimori}, {Suto}, {Suzuki}, {Takada}, {Takagi}, {Takata}, {Takita}, {Tanaka}, {Tang}, {Taranu}, {Terai}, {Toba}, {Turner}, {Uchiyama}, {Vijarnwannaluk}, {Waters}, {Yamada}, {Yamamoto}, \& {Yamashita}}]{2022PASJ...74..247A}
{Aihara}, H., {AlSayyad}, Y., {Ando}, M., {et~al.} 2022, \bibinfo{title}{{Third data release of the Hyper Suprime-Cam Subaru Strategic Program},} \pasj, 74, 247, \dodoi{10.1093/pasj/psab122}

\bibitem[{G. {Battaglia} {et~al.}(2022){Battaglia}, {Taibi}, {Thomas}, \& {Fritz}}]{2022A&A...657A..54B}
{Battaglia}, G., {Taibi}, S., {Thomas}, G.~F., \& {Fritz}, T.~K. 2022, \bibinfo{title}{{Gaia early DR3 systemic motions of Local Group dwarf galaxies and orbital properties with a massive Large Magellanic Cloud},} \aap, 657, A54, \dodoi{10.1051/0004-6361/202141528}

\bibitem[{K. {Bekki} \& M. {Chiba}(2001){Bekki} \& {Chiba}}]{2001ApJ...558..666B}
{Bekki}, K., \& {Chiba}, M. 2001, \bibinfo{title}{{Formation of the Galactic Stellar Halo. I. Structure and Kinematics},} \apj, 558, 666, \dodoi{10.1086/322300}

\bibitem[{B. {Belland} {et~al.}(2020){Belland}, {Kirby}, {Boylan-Kolchin}, \& {Wheeler}}]{2020ApJ...903...10B}
{Belland}, B., {Kirby}, E., {Boylan-Kolchin}, M., \& {Wheeler}, C. 2020, \bibinfo{title}{{NGC 6822 as a Probe of Dwarf Galactic Evolution},} \apj, 903, 10, \dodoi{10.3847/1538-4357/abb5f4}

\bibitem[{J.~S. {Bullock} \& K.~V. {Johnston}(2005){Bullock} \& {Johnston}}]{2005ApJ...635..931B}
{Bullock}, J.~S., \& {Johnston}, K.~V. 2005, \bibinfo{title}{{Tracing Galaxy Formation with Stellar Halos. I. Methods},} \apj, 635, 931, \dodoi{10.1086/497422}

\bibitem[{A. {Chiti} {et~al.}(2021){Chiti}, {Frebel}, {Simon}, {Erkal}, {Chang}, {Necib}, {Ji}, {Jerjen}, {Kim}, \& {Norris}}]{2021NatAs...5..392C}
{Chiti}, A., {Frebel}, A., {Simon}, J.~D., {et~al.} 2021, \bibinfo{title}{{An extended halo around an ancient dwarf galaxy},} Nature Astronomy, 5, 392, \dodoi{10.1038/s41550-020-01285-w}

\bibitem[{A.~P. {Cooper} {et~al.}(2010){Cooper}, {Cole}, {Frenk}, {White}, {Helly}, {Benson}, {De Lucia}, {Helmi}, {Jenkins}, {Navarro}, {Springel}, \& {Wang}}]{2010MNRAS.406..744C}
{Cooper}, A.~P., {Cole}, S., {Frenk}, C.~S., {et~al.} 2010, \bibinfo{title}{{Galactic stellar haloes in the CDM model},} \mnras, 406, 744, \dodoi{10.1111/j.1365-2966.2010.16740.x}

\bibitem[{A.~J. {Deason} {et~al.}(2022){Deason}, {Bose}, {Fattahi}, {Amorisco}, {Hellwing}, \& {Frenk}}]{2022MNRAS.511.4044D}
{Deason}, A.~J., {Bose}, S., {Fattahi}, A., {et~al.} 2022, \bibinfo{title}{{Dwarf stellar haloes: a powerful probe of small-scale galaxy formation and the nature of dark matter},} \mnras, 511, 4044, \dodoi{10.1093/mnras/stab3524}

\bibitem[{I. {Ebrov{\'a}} \& E.~L. {{\L}okas}(2015){Ebrov{\'a}} \& {{\L}okas}}]{2015ApJ...813...10E}
{Ebrov{\'a}}, I., \& {{\L}okas}, E.~L. 2015, \bibinfo{title}{{The Origin of Prolate Rotation in Dwarf Spheroidal Galaxies Formed by Mergers of Disky Dwarfs},} \apj, 813, 10, \dodoi{10.1088/0004-637X/813/1/10}

\bibitem[{C. {Filion} \& R.~F.~G. {Wyse}(2021){Filion} \& {Wyse}}]{2021ApJ...923..218F}
{Filion}, C., \& {Wyse}, R. F.~G. 2021, \bibinfo{title}{{The Far-away Blues: Exploring the Furthest Extents of the Bo{\"o}tes I Ultra-faint Dwarf Galaxy},} \apj, 923, 218, \dodoi{10.3847/1538-4357/ac2df1}

\bibitem[{D. {Foreman-Mackey} {et~al.}(2013){Foreman-Mackey}, {Hogg}, {Lang}, \& {Goodman}}]{2013PASP..125..306F}
{Foreman-Mackey}, D., {Hogg}, D.~W., {Lang}, D., \& {Goodman}, J. 2013, \bibinfo{title}{{emcee: The MCMC Hammer},} \pasp, 125, 306, \dodoi{10.1086/670067}

\bibitem[{A. {Frebel} \& J.~E. {Norris}(2015){Frebel} \& {Norris}}]{2015ARA&A..53..631F}
{Frebel}, A., \& {Norris}, J.~E. 2015, \bibinfo{title}{{Near-Field Cosmology with Extremely Metal-Poor Stars},} \araa, 53, 631, \dodoi{10.1146/annurev-astro-082214-122423}

\bibitem[{C.~S. {Frenk} {et~al.}(1988){Frenk}, {White}, {Davis}, \& {Efstathiou}}]{1988ApJ...327..507F}
{Frenk}, C.~S., {White}, S. D.~M., {Davis}, M., \& {Efstathiou}, G. 1988, \bibinfo{title}{{The Formation of Dark Halos in a Universe Dominated by Cold Dark Matter},} \apj, 327, 507, \dodoi{10.1086/166213}

\bibitem[{A. {Helmi} {et~al.}(2018){Helmi}, {Babusiaux}, {Koppelman}, {Massari}, {Veljanoski}, \& {Brown}}]{2018Natur.563...85H}
{Helmi}, A., {Babusiaux}, C., {Koppelman}, H.~H., {et~al.} 2018, \bibinfo{title}{{The merger that led to the formation of the Milky Way's inner stellar halo and thick disk},} \nat, 563, 85, \dodoi{10.1038/s41586-018-0625-x}

\bibitem[{N. {Ho} {et~al.}(2012){Ho}, {Geha}, {Munoz}, {Guhathakurta}, {Kalirai}, {Gilbert}, {Tollerud}, {Bullock}, {Beaton}, \& {Majewski}}]{2012ApJ...758..124H}
{Ho}, N., {Geha}, M., {Munoz}, R.~R., {et~al.} 2012, \bibinfo{title}{{Stellar Kinematics of the Andromeda II Dwarf Spheroidal Galaxy},} \apj, 758, 124, \dodoi{10.1088/0004-637X/758/2/124}

\bibitem[{J. {Jensen} {et~al.}(2024){Jensen}, {Hayes}, {Sestito}, {McConnachie}, {Waller}, {Smith}, {Navarro}, \& {Venn}}]{2024MNRAS.527.4209J}
{Jensen}, J., {Hayes}, C.~R., {Sestito}, F., {et~al.} 2024, \bibinfo{title}{{Small-scale stellar haloes: detecting low surface brightness features in the outskirts of Milky Way dwarf satellites},} \mnras, 527, 4209, \dodoi{10.1093/mnras/stad3322}

\bibitem[{N. {Kacharov} {et~al.}(2017){Kacharov}, {Battaglia}, {Rejkuba}, {Cole}, {Carrera}, {Fraternali}, {Wilkinson}, {Gallart}, {Irwin}, \& {Tolstoy}}]{2017MNRAS.466.2006K}
{Kacharov}, N., {Battaglia}, G., {Rejkuba}, M., {et~al.} 2017, \bibinfo{title}{{Prolate rotation and metallicity gradient in the transforming dwarf galaxy Phoenix},} \mnras, 466, 2006, \dodoi{10.1093/mnras/stw3188}

\bibitem[{E.~N. {Kirby} {et~al.}(2013){Kirby}, {Cohen}, {Guhathakurta}, {Cheng}, {Bullock}, \& {Gallazzi}}]{2013ApJ...779..102K}
{Kirby}, E.~N., {Cohen}, J.~G., {Guhathakurta}, P., {et~al.} 2013, \bibinfo{title}{{The Universal Stellar Mass-Stellar Metallicity Relation for Dwarf Galaxies},} \apj, 779, 102, \dodoi{10.1088/0004-637X/779/2/102}

\bibitem[{E.~L. {{\L}okas} {et~al.}(2013){{\L}okas}, {Gajda}, \& {Kazantzidis}}]{2013MNRAS.433..878L}
{{\L}okas}, E.~L., {Gajda}, G., \& {Kazantzidis}, S. 2013, \bibinfo{title}{{Tidal tails of dwarf galaxies on different orbits around the Milky Way},} \mnras, 433, 878, \dodoi{10.1093/mnras/stt774}

\bibitem[{K. {Malhan} {et~al.}(2021){Malhan}, {Yuan}, {Ibata}, {Arentsen}, {Bellazzini}, \& {Martin}}]{2021ApJ...920...51M}
{Malhan}, K., {Yuan}, Z., {Ibata}, R.~A., {et~al.} 2021, \bibinfo{title}{{Evidence of a Dwarf Galaxy Stream Populating the Inner Milky Way Halo},} \apj, 920, 51, \dodoi{10.3847/1538-4357/ac1675}

\bibitem[{N.~F. {Martin} {et~al.}(2022){Martin}, {Venn}, {Aguado}, {Starkenburg}, {Gonz{\'a}lez Hern{\'a}ndez}, {Ibata}, {Bonifacio}, {Caffau}, {Sestito}, {Arentsen}, {Allende Prieto}, {Carlberg}, {Fabbro}, {Fouesneau}, {Hill}, {Jablonka}, {Kordopatis}, {Lardo}, {Malhan}, {Mashonkina}, {McConnachie}, {Navarro}, {S{\'a}nchez-Janssen}, {Thomas}, {Yuan}, \& {Mucciarelli}}]{2022Natur.601...45M}
{Martin}, N.~F., {Venn}, K.~A., {Aguado}, D.~S., {et~al.} 2022, \bibinfo{title}{{A stellar stream remnant of a globular cluster below the metallicity floor},} \nat, 601, 45, \dodoi{10.1038/s41586-021-04162-2}

\bibitem[{K.~J. {Mighell} \& C.~J. {Burke}(1999){Mighell} \& {Burke}}]{1999AJ....118..366M}
{Mighell}, K.~J., \& {Burke}, C.~J. 1999, \bibinfo{title}{{WFPC2 Observations of the Ursa Minor Dwarf Spheroidal Galaxy},} \aj, 118, 366, \dodoi{10.1086/300923}

\bibitem[{T. {Miyoshi} \& M. {Chiba}(2020){Miyoshi} \& {Chiba}}]{2020ApJ...905..109M}
{Miyoshi}, T., \& {Chiba}, M. 2020, \bibinfo{title}{{Long-term Orbital Evolution of Galactic Satellites and the Effects on Their Star Formation Histories},} \apj, 905, 109, \dodoi{10.3847/1538-4357/abc486}

\bibitem[{I. {Ogami} {et~al.}(2024){Ogami}, {Komiyama}, {Chiba}, {Tanaka}, {Guhathakurta}, {Kirby}, {Wyse}, {Filion}, {Kirihara}, {Ishigaki}, \& {Hayashi}}]{2024ApJ...971..107O}
{Ogami}, I., {Komiyama}, Y., {Chiba}, M., {et~al.} 2024, \bibinfo{title}{{Detection of a Spatially Extended Stellar Population in M33: A Shallow Stellar Halo?},} \apj, 971, 107, \dodoi{10.3847/1538-4357/ad5445}

\bibitem[{S. {Okamoto} {et~al.}(2019){Okamoto}, {Arimoto}, {Ferguson}, {Irwin}, {Bernard}, \& {Utsumi}}]{2019ApJ...884..128O}
{Okamoto}, S., {Arimoto}, N., {Ferguson}, A. M.~N., {et~al.} 2019, \bibinfo{title}{{Stellar Population and Structural Properties of Dwarf Galaxies and Young Stellar Systems in the M81 Group},} \apj, 884, 128, \dodoi{10.3847/1538-4357/ab44a7}

\bibitem[{A.~B. {Pace} {et~al.}(2020){Pace}, {Kaplinghat}, {Kirby}, {Simon}, {Tollerud}, {Mu{\~n}oz}, {C{\^o}t{\'e}}, {Djorgovski}, \& {Geha}}]{2020MNRAS.495.3022P}
{Pace}, A.~B., {Kaplinghat}, M., {Kirby}, E., {et~al.} 2020, \bibinfo{title}{{Multiple chemodynamic stellar populations of the Ursa Minor dwarf spheroidal galaxy},} \mnras, 495, 3022, \dodoi{10.1093/mnras/staa1419}

\bibitem[{C. {Palma} {et~al.}(2003){Palma}, {Majewski}, {Siegel}, {Patterson}, {Ostheimer}, \& {Link}}]{2003AJ....125.1352P}
{Palma}, C., {Majewski}, S.~R., {Siegel}, M.~H., {et~al.} 2003, \bibinfo{title}{{Exploring Halo Substructure with Giant Stars. IV. The Extended Structure of the Ursa Minor Dwarf Spheroidal Galaxy},} \aj, 125, 1352, \dodoi{10.1086/367594}

\bibitem[{J. {Pe{\~n}arrubia} {et~al.}(2008){Pe{\~n}arrubia}, {Navarro}, \& {McConnachie}}]{2008ApJ...673..226P}
{Pe{\~n}arrubia}, J., {Navarro}, J.~F., \& {McConnachie}, A.~W. 2008, \bibinfo{title}{{The Tidal Evolution of Local Group Dwarf Spheroidals},} \apj, 673, 226, \dodoi{10.1086/523686}

\bibitem[{L. {Querci} {et~al.}(2025){Querci}, {Pallottini}, {Branca}, \& {Salvadori}}]{2025A&A...694A..17Q}
{Querci}, L., {Pallottini}, A., {Branca}, L., \& {Salvadori}, S. 2025, \bibinfo{title}{{Stellar halos tracing the assembly of ultra-faint dwarf galaxies},} \aap, 694, A17, \dodoi{10.1051/0004-6361/202452476}

\bibitem[{K.~S. Sato {et~al.}(2025)Sato, Komiyama, Okamoto, Yagi, Ogami, Tanaka, Arimoto, Chiba, Kirby, \& Wyse}]{sato2025starformationchemicalevolution}
Sato, K.~S., Komiyama, Y., Okamoto, S., {et~al.} 2025, \bibinfo{title}{The Star Formation and Chemical Evolution Histories of Ursa Minor Dwarf Spheroidal Galaxy,} \doarXiv{2505.13161}

\bibitem[{E.~F. {Schlafly} \& D.~P. {Finkbeiner}(2011){Schlafly} \& {Finkbeiner}}]{2011ApJ...737..103S}
{Schlafly}, E.~F., \& {Finkbeiner}, D.~P. 2011, \bibinfo{title}{{Measuring Reddening with Sloan Digital Sky Survey Stellar Spectra and Recalibrating SFD},} \apj, 737, 103, \dodoi{10.1088/0004-637X/737/2/103}

\bibitem[{L. {Searle} \& R. {Zinn}(1978){Searle} \& {Zinn}}]{1978ApJ...225..357S}
{Searle}, L., \& {Zinn}, R. 1978, \bibinfo{title}{{Composition of halo clusters and the formation of the galactic halo.},} \apj, 225, 357, \dodoi{10.1086/156499}

\bibitem[{F. {Sestito} {et~al.}(2023){Sestito}, {Zaremba}, {Venn}, {D'Aoust}, {Hayes}, {Jensen}, {Navarro}, {Jablonka}, {Fern{\'a}ndez-Alvar}, {Glover}, {McConnachie}, \& {Chen{\'e}}}]{2023MNRAS.525.2875S}
{Sestito}, F., {Zaremba}, D., {Venn}, K.~A., {et~al.} 2023, \bibinfo{title}{{The extended 'stellar halo' of the Ursa Minor dwarf galaxy},} \mnras, 525, 2875, \dodoi{10.1093/mnras/stad2427}

\bibitem[{Y. {Tarumi} {et~al.}(2021){Tarumi}, {Yoshida}, \& {Frebel}}]{2021ApJ...914L..10T}
{Tarumi}, Y., {Yoshida}, N., \& {Frebel}, A. 2021, \bibinfo{title}{{Formation of an Extended Stellar Halo around an Ultra-faint Dwarf Galaxy Following One of the Earliest Mergers from Galactic Building Blocks},} \apjl, 914, L10, \dodoi{10.3847/2041-8213/ac024e}

\bibitem[{J. {Wolf} {et~al.}(2010){Wolf}, {Martinez}, {Bullock}, {Kaplinghat}, {Geha}, {Mu{\~n}oz}, {Simon}, \& {Avedo}}]{2010MNRAS.406.1220W}
{Wolf}, J., {Martinez}, G.~D., {Bullock}, J.~S., {et~al.} 2010, \bibinfo{title}{{Accurate masses for dispersion-supported galaxies},} \mnras, 406, 1220, \dodoi{10.1111/j.1365-2966.2010.16753.x}

\end{thebibliography}
\bibliographystyle{aasjournalv7.bst}
\end{document}